\documentclass[twocolumn,a4paper,aps,pra,longbibliography]{revtex4-1}
\usepackage[T1]{fontenc}
\usepackage[ngerman,english]{babel}
\usepackage[utf8]{inputenc}
\usepackage{graphicx}
\usepackage{float}
\usepackage{array}
\usepackage{braket}
\usepackage[colorlinks=true,urlcolor = blue,linkcolor=blue,citecolor=blue]{hyperref}
\usepackage[babel,german=quotes]{csquotes}
\usepackage{mathtools}
\usepackage{bbold}
\usepackage{xcolor}
\usepackage [caption=false]{subfig}
\usepackage{hyperref}

\usepackage{tikz}
\usetikzlibrary{3d,positioning,arrows,decorations.markings,shapes.geometric}
\usetikzlibrary{external}
\tikzsetexternalprefix{tikzexternal/}
\tikzset{external/system call={lualatex -shell-escape -halt-on-error
-interaction=batchmode -jobname "\image" "\texsource"}}
\usepackage{pgfplots}
\pgfplotsset{compat=1.7}
\pgfplotsset{colormap={redyellow}{rgb255(0cm)=(255,0,0); rgb255(1cm)=(255,255,0)}}
\pgfplotsset{colormap={greenyellow}{rgb255(0cm)=(0,128,0); rgb255(1cm)=(255,255,0)}}
\pgfplotsset{colormap={cool}{rgb255(0cm)=(255,255,255); rgb255(1cm)=(0,128,255); rgb255(2cm)=(255,0,255)}}
\pgfplotsset{colormap={bluered}{rgb255(0cm)=(0,0,180); rgb255(1cm)=(0,255,255); rgb255(2cm)=(100,255,0);rgb255(3cm)=(255,255,0); rgb255(4cm)=(255,0,0); rgb255(5cm)=(128,0,0)}}
\pgfplotsset{/pgfplots/colormap={hot2}{[1cm]rgb255(0cm)=(0,0,0)rgb255(3cm)=(255,0,0)rgb255(6cm)=(255,255,0)rgb255(8cm)=(255,255,255)}}
\pgfplotsset{/pgfplots/colormap={winter}{rgb255=(0,0,255) rgb255=(0,255,128)}}
\pgfplotsset{invoke before crossref tikzpicture={\tikzexternaldisable},invoke after crossref tikzpicture={\tikzexternalenable}}
\tikzset{%/pgf/number format/use comma, 
/pgf/number format/1000 sep={}}
\definecolor{cset-aps-red}{RGB}{219,0,43}

\usepackage{cancel}

\begin{document}

%\date{\today}
\author{Fabio Di Pumpo}
\email{fabio.di-pumpo@uni-ulm.de}
\author{Matthias Freyberger}
%\email{matthias.freyberger@uni-ulm.de}
\affiliation{Institut für Quantenphysik and Center for Integrated Quantum Science and Technology ($\text{IQ}^{\text{ST}}$), Universität Ulm, Albert Einstein Allee 11, D-89069 Ulm, Germany}
\title{Pointer-based model for state reduction in momentum space\\
\normalfont{\normalsize{Published in \href{https://doi.org/10.1140/epjd/e2019-100226-1}{Eur. Phys. J. D \textbf{73}, 163 (2019)}}}}
%\email{matthias.freyberger@uni-ulm.de}

\begin{abstract}
	We revisit the pointer-based measurement concept of von Neumann which allows us to model a quantum counterpart of the classical time-of-flight (ToF) momentum. Our approach is based on the Hamiltonian for a particle interacting with two quantum pointers serving as basic measurement devices. The corresponding dynamics leads to a pointer-based ToF observable for the operational momentum of the particle. We can consider single measurements of our quantum pointers and show that this process will result in a state reduction for a single particle being downstream of the time-of-flight setup.
\end{abstract}
\maketitle

\section{Introduction}

The quantum formalism \cite{VonNeumann1968MathematischeGrundlagen,gallone2014hilbert,david2014formalisms} has outstanding efficiency and elegance: a small set of axiomatic rules explains the vast majority of known experimental data.

In contrast to this enormous success some basic operational concepts, which are straightforward in classical physics, are by far not so obvious in quantum mechanics \cite{lamb1969operational,PhysRevD.24.1516,PhysRevD.26.1862,busch1997operational,klyshko1998basic,dariano2007Operational}. Using well-established mathematics we can extract classical information from a quantum system. It is a standard technique to represent the quantum state in a basis of eigenstates corresponding to a certain observable and hence to specify the statistics of the measurable values. 

The other approach is operationally less known and typically much more complicated \cite{klyshko1998basic}. If we measure a set of classical values, for example positions and momenta \cite{PhysRevA.51.R2661}, what will these data reveal about the underlying quantum system? This question becomes quite obvious in the conception of a classical state versus a quantum state. For the classical state of a particle we measure a single point in phase space, which has a simple operational meaning. In the quantum case we have to reconstruct a complete phase-space distribution \cite{HILLERY1984121,schleich2001quantum}, like the Wigner function, from infinitely many quadrature distributions \cite{vogel1989determination,leonhardt1997measuring,paris2004quantum}.

Closely related to this topic is the problem of simultaneously measuring conjugate observables \cite{BLTJ:BLTJ1684,park_simultaneous_1968,STENHOLM1992233}. Classically, the simultaneous measurement of position and momentum is operationally simple and in fact leads to the aforementioned classical state of a particle. In quantum mechanics, one can formulate various operational scenarios which all mean a certain kind of simultaneous measurement, but with different results and, for example, very different uncertainty relations \cite{busch1985indeterminacy,WODKIEWICZ1987207,busshardt2010timing,heese2013entropic}. 

A similar situation can be found for the time-of-arrival problem \cite{muga2000arrival}, which also has a straightforward answer in classical mechanics. However, in the quantum case \cite{grot1996time,delgado1997arrival,leon_time--arrival_1997,anastopoulos_time--arrival_2006,anastopoulos_time--arrival_2012,anastopoulos_time--arrival_2017,anastopoulos_time_2019} it becomes a complicated foundational issue, questioning the role of time in the theory \cite{aharonov_time1961,kijowski_time1974,peres_measurement1980,aharonov_measurement1998,yearsley_quantum2011}. In fact one needs several operationally defined time-of-arrival operators.

In the present contribution we ask a question touching this time-of-arrival problem: How can we operationally define the momentum of a non-relativistic quantum particle? Our answer cannot be that we have to decompose the corresponding quantum state in terms of momentum eigenstates from which one is selected in a ad hoc state-reduction process. We rather start from the naive operational concept of momentum in classical physics, basically set up with time-of-flight measurements. We transfer them to the quantum realm as closely as possible using two pointer systems \cite{VonNeumann1968MathematischeGrundlagen,PhysRevD.24.1516} coupled to the particle as proposed by the von Neumann model \cite{VonNeumann1968MathematischeGrundlagen}. By analyzing the joint dynamics of these three systems we can define a ToF observable, which allows us to describe the particle in momentum space without referring to the abstract momentum operator. Actually, we expect that by simply reading off the pointers we assign a specific momentum to the particle, that is we obtain a reduction of the corresponding wave packet in momentum space.

The work is organized as follows: In Sect.\,\ref{sec:ToFMes} we define the ToF operator, motivated by a classical time-of-flight concept \cite{muga2000arrival}. To achieve this operational quantity, we construct the full three particle Hamiltonian including the interaction between pointers and system and solve the corresponding dynamics in the Heisenberg picture. We then discuss in what sense the expectation value of this ToF observable reveals meaningful information about the mean momentum of the particle. In Sect.\,\ref{sec:DynSingleMom} we analyze a single ToF measurement value and answer the question if a single position measurement on each pointer defines a reasonable momentum of the particle and hence leads to a state narrowed down in momentum space. For this purpose we change to the Schrödinger picture and calculate the conditioned post-measurement momentum and the corresponding variance to see the mentioned state-reduction process. In Sect.\,\ref{sec:Conclusion} we summarize our results and give an outlook for further studies based on the presented model.

\section{Time-of-flight measurements}
\label{sec:ToFMes}
\subsection{Classical versus quantum time-of-flight measurements}

\begin{figure}[h]
\centering	
\subfloat[Classical ToF\hspace*{0em}]
{\includegraphics[width=0.4\textwidth]{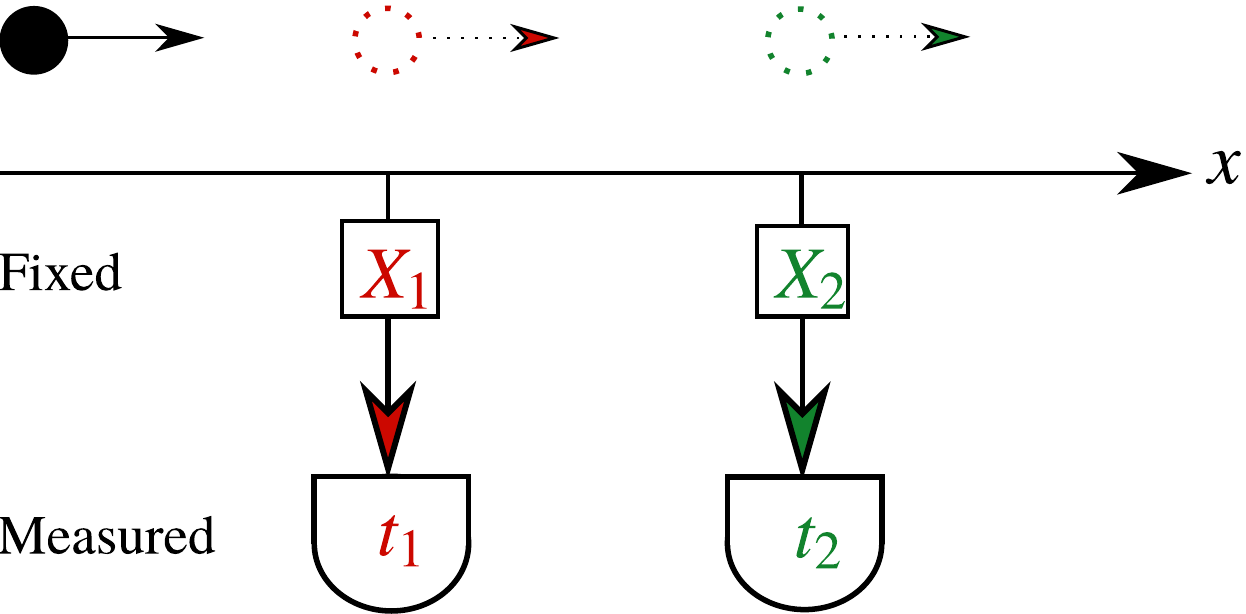}}

\subfloat[Quantum ToF\hspace*{0em}]
{\includegraphics[width=0.4\textwidth]{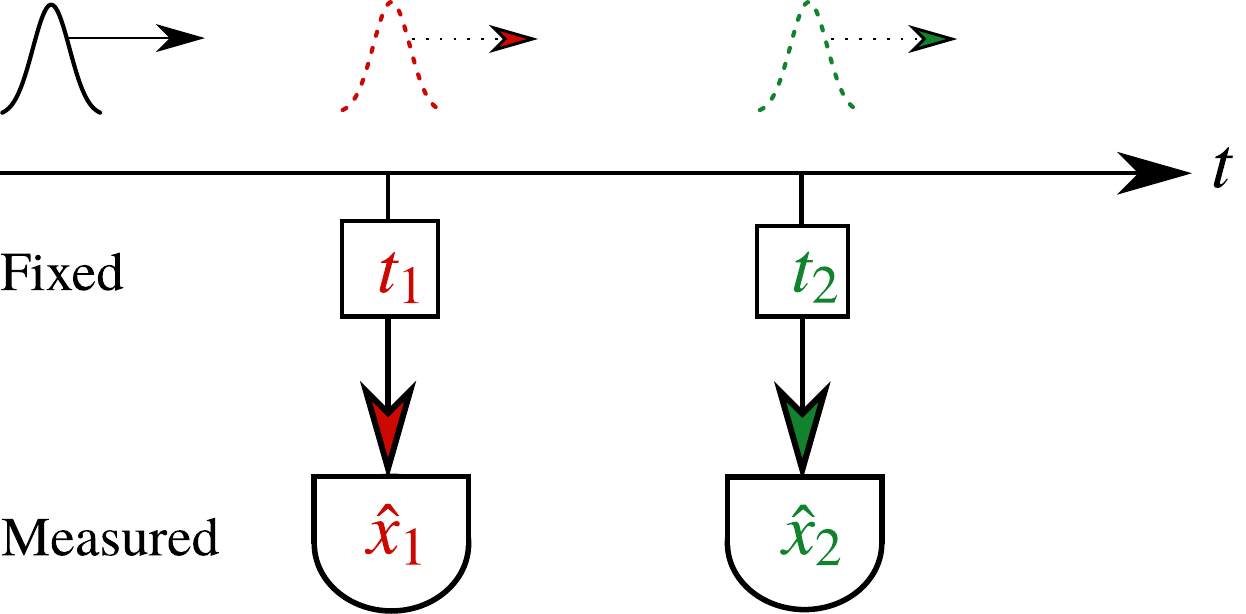}}
	\caption[Classical and quantum time-of-flight measurement]{The upper scheme (a) illustrates the naive classical time-of-flight momentum measurement. The mean momentum $\mathcal{P}$ of a particle is calculated from the measurement of the times at the fixed positions $X_1$ and $X_2$ according to Eq.\,(\ref{eq:ClassTimeOfFlight}). In the lower scheme (b), the particle is in a quantum state, represented by a wave packet. Moreover, we also assume the detectors to be quantum devices. Hence, we can no longer fix their positions but rather define times $t_1$ and $t_2$ for which probabilistic positions are registered. As a consequence, we obtain a meaningful ToF observable for the momentum $\hat{\mathcal{P}}$, Eq.\,(\ref{eq:QTimeOfFlight}), of our particle.}
\label{fig:ClassAndQuantToFTotal}
\end{figure}

Our aim is to model the measurement for the momentum of a free quantum particle. In classical mechanics such measurements are based on simple \textit{time-of-flight} (ToF) concepts, see Fig.\,\hyperref[fig:ClassAndQuantToFTotal]{1a}: We fix two detectors at positions $X_1$ and $X_2>X_1$. For a particle of mass $M$ we then obtain a mean momentum 
\begin{equation}
\mathcal{P}=M\frac{X_2-X_1}{t_2-t_1}
\label{eq:ClassTimeOfFlight}
\end{equation}
where $t_1$ and $t_2$ are the times measured when the particle passes the positions $X_1$ and $X_2$. Obviously, in this naive classical case a time difference is measured while the positions of the detectors stay fixed.

Starting from this naive classical scheme, we have to interchange the role of time and position in a possible ToF scheme in the quantum case, see Fig.\,\hyperref[fig:ClassAndQuantToFTotal]{1b}: Now we fix two times $t_2>t_1$ at which we determine the probabilistic positions of the quantum particle which is now described by a wave packet. Moreover, in the quantum case we will also include the detectors themselves as simplified quantum devices \footnote{This quantum scenario of a ToF measurement is closest to the classical scenario of taking snapshots on a movie with a camera of the classical particle traveling along the $x$-axis.}.
%\vspace{-0.55em}
Despite such an exchange of measurement principles, we can still see the analogy between classical and quantum time-of-flight measurements. These considerations allow us to formulate the quantum counterpart of a mean momentum, 
\begin{equation}
\hat{\mathcal{P}}=M\frac{\hat{x}_2-\hat{x}_1}{t_2-t_1},
\label{eq:QTimeOfFlight}
\end{equation}
in which the operators $\hat{x}_i$ will describe what we can read off the measurement devices. In what follows, we will develop an operational definition of these position-like operators.

\subsection{Quantum pointers}

As already mentioned, we will include measurement devices as additional quantum systems, here called \textit{\smash{quantum} pointers}. We will keep these pointers as simple as possible. They just have to be coupled to the particle if we want them to record any measurement data.

A basic concept goes back to von Neumann \cite{VonNeumann1968MathematischeGrundlagen}. The idea is that an observable $\hat{A}=\hat{A}^{\dagger}$ of the particle to be measured unitarily displaces the position of the pointer. The corresponding displacement operator then reads
\begin{equation}
\hat{T}_{\hat{A}}=\exp{\left(-\frac{i}{\hbar}\kappa\,\hat{A}\otimes\hat{p}\right)},
\label{eq:InterHam1}
\end{equation}
where $\kappa$ is a constant determining the coupling strength and $\hat{p}$ denotes the momentum operator of the quantum pointer.

In particular, if we choose the position operator $\hat{A}=\hat{X}$ of the particle in Eq.\,(\ref{eq:InterHam1}) and the operator acts on the states $\ket{\Phi}$ and $\ket{\psi}$ of system and pointer, respectively, we find the displacement 
\begin{equation}
\left(\bra{X}\otimes\bra{x}\right)\hat{T}_{\hat{X}}\left(\ket{\Phi}\otimes\ket{\psi}\right)=\Phi\left(X\right)\cdot\psi\left(x-\kappa\,X\right)
\label{eq:InterHam2}
\end{equation}
in position representation using the eigenstates $\ket{X}$ and $\ket{x}$. Thus, we obtain a displaced probability density\\ $\left|\psi\left(x-\kappa\,X\right)\right|^2$ of the pointer which depends on the position eigenvalue $X$ of the particle. This displaced probability density allows us to obtain information about the statistics of the position observable $\hat{X}$ of the particle by measuring the position observable $\hat{x}$ of the pointer. Clearly, we have to know the initial state $\ket{\psi}$ of the pointer.

A simplified interaction Hamiltonian that leads to the unitary operator in Eq.\,(\ref{eq:InterHam1}) is of the form
\begin{equation}
\hat{H}_{\mathrm{int}}\left(t\right)=f\left(t\right)\hat{A}\otimes\hat{p},
\label{eq:InterHam3}
\end{equation}
where $f\left(t\right)$ is a time-dependent \textit{coupling function} which will be specified later for our particular ToF scheme.

\subsection{ToF-Hamiltonian and scaling}

Having established the basic form of an interaction Hamiltonian, Eq.\,(\ref{eq:InterHam3}), between a single pointer and the particle, we now formulate the total Hamiltonian of our ToF model which requires two pointers, denoted by index $i=1,2$. First we have the free evolution 
\begin{equation}
\hat{H}_0=\frac{\hat{P}^2}{2M}+\frac{{\hat{p}^2}_1}{2m_1}+\frac{{\hat{p}^2}_2}{2m_2} 
\label{eq:InterHam4}
\end{equation}
of all subsystems, where $\hat{P}$ and $\hat{p}_i$ are the momentum operators of the particle with mass $M$ and the pointers with masses $m_i$ for $i=1,2$, respectively. Additionally, two time-dependent interaction terms of the form of Eq.\,(\ref{eq:InterHam3}) between the particle and both pointers appear in
\begin{equation}
\hat{H}_{\mathrm{ToF}}\left(t\right)=f_1\left(t\right)\hat{X}\otimes\hat{p}_1+f_2\left(t\right)\hat{X}\otimes\hat{p}_2, 
\label{eq:InterHam5}
\end{equation}
with the coupling functions $f_i\left(t\right)$ and $\hat{A}=\hat{X}$ to model the ToF idea of Fig.\,\hyperref[fig:ClassAndQuantToFTotal]{1b}. That is, the functions $f_i\left(t\right)$ will produce an appropriate time-ordered coupling of both pointers to the particle. This result leads to the total three-particle ToF-Hamiltonian
\begin{equation}
\hat{H}\left(t\right)=\hat{H}_0+\hat{H}_{\mathrm{ToF}}\left(t\right).
\label{eq:Hamiltonian1}
\end{equation}
Before solving the corresponding dynamics, we define dimensionless observables 
\begin{subequations}
\begin{equation}
\hat{X}'\coloneqq\frac{\hat{X}}{a},\,\hat{P}'\coloneqq\frac{a}{\hbar}\hat{P},\,\hat{x}'_i\coloneqq\frac{\hat{x}_i}{a}\quad\mathrm{and}\quad\hat{p}'_i\coloneqq\frac{a}{\hbar}\hat{p}_i,
\label{eq:Rescale1}
\end{equation}
where the constant $a$ denotes a length scale which can be chosen to be the initial width of the particle's wave packet. Besides, we also use a dimensionless time $t'$ given by
\begin{equation}
t'\coloneqq \frac{\hbar}{M a^2}\,t,
\label{eq:Rescale2}
\end{equation}
\end{subequations}
in which $Ma^2/\hbar$ represents a typical spreading time of a wave packet over the length scale $a$.

\subsection{Dynamics}

In order to find a suitable definition for our ToF observable, Eq.\,(\ref{eq:QTimeOfFlight}), we now evaluate the dynamics given by $\hat{H}\left(t\right)$, Eq.\,(\ref{eq:Hamiltonian1}), in the Heisenberg picture. The corresponding equation of motion
\begin{equation}
\frac{\mathrm{d}}{\mathrm{d}t}\hat{A}^{\left(H\right)}\left(t\right)=\frac{i}{\hbar}\left[\hat{H}^{\left(H\right)}\left(t\right),\hat{A}^{\left(H\right)}\left(t\right)\right]
\label{eq:HeisenbergMotion1}
\end{equation}
for an Heisenberg operator $\hat{A}^{\left(H\right)}\left(t\right)$ will be written in the aforementioned scaled variables where from now on we omit the prime for clarity. We then find the specific set of equations
\begin{subequations}
\begin{align}
\frac{\mathrm{d}}{\mathrm{d}t}\hat{p}_i\left(t\right)&=0,  \\
\frac{\mathrm{d}}{\mathrm{d}t}\hat{P}\left(t\right)&=-f_1\left(t\right)\hat{p}_1\left(t\right)-f_2\left(t\right)\hat{p}_2\left(t\right), \\
\frac{\mathrm{d}}{\mathrm{d}t}\hat{X}\left(t\right)&=\hat{P}\left(t\right) 
\end{align}
and
\begin{equation}
\frac{\mathrm{d}}{\mathrm{d}t}\hat{x}_i\left(t\right)=f_i\left(t\right)\hat{X}\left(t\right)+\frac{M}{m_i}\hat{p}_i\left(t\right),
\end{equation}
\end{subequations}
where we have also suppressed the superscripts related to the different pictures, since we solely consider the Heisenberg picture. This set of equations can be solved in a straightforward manner. The momenta of the pointers
\begin{equation}
\hat{p}_i\left(t\right)=\hat{p}_i\left(0\right) \label{eq:HeisenbergMotion10} 
\end{equation}
are constants of motion. Then, we find the observables
\begin{equation}
\hat{P}\left(t\right)=\hat{P}\left(0\right)-a_1\left(t\right)\hat{p}_1\left(0\right)-a_2\left(t\right)\hat{p}_2\left(0\right), \label{eq:HeisenbergMotion11} 
\end{equation}
and
\begin{equation}
\hat{X}\left(t\right)=\hat{X}\left(0\right)+\hat{P}\left(0\right)t-b_1\left(t\right)\hat{p}_1\left(0\right)-b_2\left(t\right)\hat{p}_2\left(0\right) 
\label{eq:HeisenbergMotion12} 
\end{equation}
of the particle, where we have introduced the time-dependent coefficients
\begin{equation}
a_i\left(t\right)\coloneqq\int_0^{t}{\mathrm{d}\tau\,f_i(\tau)}
\label{eq:CouplingInt1}
\end{equation}
and 
\begin{equation}
b_i\left(t\right)\coloneqq\int_0^{t}{\mathrm{d}\tau\,a_i(\tau)}.
\label{eq:CouplingInt2}
\end{equation}
For the position observables of the pointers we arrive at a slightly more involved expression
\begin{align}
\hat{x}_i\left(t\right)=&\hat{x}_i\left(0\right)+\frac{M}{m_i}\hat{p}_i\left(0\right)t \nonumber \\
&+a_i\left(t\right)\hat{X}\left(0\right)+\hat{P}\left(0\right)\int_0^{t}{\mathrm{d}\tau\,\tau f_i(\tau)} \nonumber \\
&-\hat{p}_1\left(0\right)\int_0^{t}{\mathrm{d}\tau\,f_i(\tau)b_1(\tau)} \nonumber \\
&-\hat{p}_2\left(0\right)\int_0^{t}{\mathrm{d}\tau\,f_i(\tau)b_2(\tau)}. 
\label{eq:HeisenbergMotion13} 
\end{align}

We can clearly see that the pointers now contain information on the particle's position $\hat{X}\left(0\right)$ and momentum $\hat{P}\left(0\right)$. However, also the pointers influence each other in a time-ordered manner, which becomes apparent in the coupling integrals over the products $f_i(\tau)b_j(\tau)$ in Eq.\,(\ref{eq:HeisenbergMotion13}).

In order to solve the integrals in Eqs.\,(\ref{eq:CouplingInt1}) and (\ref{eq:CouplingInt2}) explicitly, we now choose coupling functions $f_i\left(t\right)$ which allow us to perform the integrations in Eq.\,(\ref{eq:HeisenbergMotion13}). We can get the idea by choosing
\begin{equation}
f_i\left(t\right)=\kappa\,\delta\left(t-t_i\right),
\label{eq:Coupling1}
\end{equation}
where at time $t_i$ the coupling acts instantaneously with the coupling constant $\kappa$ already introduced in Eq.\,(\ref{eq:InterHam1}). Furthermore, to realize a time-of-flight measurement the second coupling has to act after the first one. Finally, we read off the pointers at time $T$. Hence we require the time sequence $T>t_2 >t_1> 0$.

With this choice, we can evaluate the observables at the readout time $T$. For the particle we find
\begin{equation}
\hat{P}\left(T\right)=\hat{P}\left(0\right)-\kappa\left[\hat{p}_1\left(0\right)+\hat{p}_2\left(0\right)\right] \label{eq:HeisenbergMotion15} 
\end{equation}
and
\begin{align}
\hat{X}\left(T\right)=&\hat{X}\left(0\right)+\hat{P}\left(0\right)T \nonumber \\
&-\kappa\left[\hat{p}_1\left(0\right)\left(T-t_1\right)+\hat{p}_2\left(0\right)\left(T-t_2\right)\right], \label{eq:HeisenbergMotion16}
\end{align} 
while we obtain for the position observables of the pointers
\begin{equation}
\hat{x}_1\left(T\right)=\hat{x}_1\left(0\right)+\frac{M}{m_1}\hat{p}_1\left(0\right)T+\kappa\,\hat{X}\left(0\right)+\kappa\,\hat{P}\left(0\right)t_1 \label{eq:HeisenbergMotion17} 
\end{equation}
and 
\begin{align}
\hat{x}_2\left(T\right)=&\hat{x}_2\left(0\right)+\frac{M}{m_2}\hat{p}_2\left(0\right)T+\kappa\,\hat{X}\left(0\right)+\kappa\,\hat{P}\left(0\right)t_2 \nonumber \\
&-\kappa^2\hat{p}_1\left(0\right)\left(t_2-t_1\right). \label{eq:HeisenbergMotion18}
\end{align}
The second pointer experiences a perturbation from the first pointer because both pointers are coupled via the particle. Therefore, the time ordering of the quantum ToF concept becomes important.

It is noteworthy that we can measure the observables $\hat{x}_1\left(T\right)$ and $\hat{x}_2\left(T\right)$ simultaneously because they commute.

\subsection{ToF observable}

Based on this result we can now define a \textit{ToF observable} in analogy to the classical time-of-flight measurement, Eq.\,(\ref{eq:ClassTimeOfFlight}), for a free particle. In dimensionless form this observable reads
\begin{equation}
\hat{\mathcal{P}}\coloneqq\frac{\hat{x}_2\left(T\right)-\hat{x}_1\left(T\right)}{t_2-t_1}.
\label{eq:OperationalMomentum}
\end{equation}
Inserting the time-evolved position observables of the pointers, Eqs.\,(\ref{eq:HeisenbergMotion17}) and (\ref{eq:HeisenbergMotion18}), in the ToF observable yields
\begin{align}
\hat{\mathcal{P}}=&\kappa\,\hat{P}\left(0\right)+\frac{\hat{x}_2\left(0\right)-\hat{x}_1\left(0\right)}{t_2-t_1} \nonumber \\
&-\kappa^2\hat{p}_1\left(0\right)+\frac{T M}{t_2-t_1}\left(\frac{\hat{p}_2\left(0\right)}{m_2}-\frac{\hat{p}_1\left(0\right)}{m_1}\right). 
\label{eq:Operational2}
\end{align}
Furthermore, we can always prepare pointers for which the mean initial positions and momenta vanish, that is 
\begin{equation}
\left<\hat{x}_i\left(0\right)\right>=\left<\hat{p}_i\left(0\right)\right>=0.
\label{eq:CondInitialExp}
\end{equation}
With this assumption, we obtain the expectation value
\begin{equation}
\left<\hat{\mathcal{P}}\right>=\kappa\left<\hat{P}\left(0\right)\right>
\label{eq:Moment3}
\end{equation}
for our ToF observable defined in Eq.\,(\ref{eq:OperationalMomentum}). Hence our pointer-based model for a ToF measurement fulfills a basic requirement: Averaging over many values of the ToF observable, Eq.\,(\ref{eq:OperationalMomentum}), delivers the mean momentum of the particle's initial wave packet.

\section{Operational state reduction in momentum space}
\label{sec:DynSingleMom}
\subsection{Momentum via ToF}

So far, we have considered statistical quantities obtained from many measurements. But what happens after a single ToF measurement? That is, we only register two pointer positions $x_i$ in a single measurement run. Regarding Eqs.\,(\ref{eq:OperationalMomentum}) and (\ref{eq:Moment3}) we have to assign the \textit{\smash{ToF value}}
\begin{equation}
P_{\mathrm{out}}\coloneqq\frac{1}{\kappa}\frac{x_2-x_1}{t_2-t_1}
\label{eq:OutcomeMomentum}
\end{equation}
to the outgoing particle. To further examine this single-run scenario, we change to the Schrödinger picture and derive the state of the total system after both couplings and after a specific measurement of the pointers. This picture allows us to analyze in what sense a single value of our ToF quantity, Eq.\,(\ref{eq:OutcomeMomentum}), defines a \textit{\smash{meaningful momentum}} of the particle being downstream of the ToF measurement. Our expectation is that reading off the pointers and attaching a certain value $P_{\mathrm{out}}$ to the particle shows two effects: (i) The conditioned wave packet of the particle being measured will be centered around the assigned momentum $P_{\mathrm{out}}$ and (ii) the corresponding width will be narrowed down compared to the initial width before any measurement. We examine both points in the following paragraph.

\subsection{Initial states and time evolution}

For the initial state of the particle we choose the real-valued Gaussian
\begin{align}
\left<P|\Phi\right>=\widetilde{\Phi}\left(P\right)=\frac{\exp{\left(-\frac{1}{4}\frac{\left(P-P_0\right)^2}{\Delta P_0^2}\right)}}{\left(2\pi\Delta P_0^2\right)^{1/4}}  
\label{eq:InitialGaussianParticle}
\end{align}
in momentum representation, in which $P_0$ is the initial mean momentum and $\Delta P_0^2$ the corresponding variance. For simplicity,  we assume the particle to be prepared such that the initial mean position vanishes. 

The pointers are initially described by separable, real-valued Gaussians 
\begin{align}
\psi_i\left(x_i\right)=\frac{\exp{\left(-\frac{1}{4}\frac{x_i^2}{\Delta x_i^2}\right)}}{\left(2\pi\Delta x_i^2\right)^{1/4}} 
\label{eq:RealValuedGaussianParticle}
\end{align}
in position representation obeying Eq.\,(\ref{eq:CondInitialExp}) for $i=1,2$. We note that these pointer states, Eq.\,(\ref{eq:RealValuedGaussianParticle}), represent minimum uncertainty states, for which the equality $\Delta x_i^2\Delta p_i^2=1/4$ holds.

With the ToF Hamiltonian, Eq.\,(\ref{eq:Hamiltonian1}), and the chosen scaling, Eqs.\,(\ref{eq:Rescale1}) and (\ref{eq:Rescale2}), we propagate the initial wave function 
\begin{equation}
\Psi\left(P,x_1,x_2,0\right)=\widetilde{\Phi}\left(P\right)\cdot\psi_1\left(x_1\right)\cdot\psi_2\left(x_2\right)
\label{eq:TotalIniStateParticlePointer}
\end{equation}
to find the time-evolved state
\begin{equation}
\Psi\left(P,x_1,x_2,T\right)=\left(\bra{P}\otimes\bra{x_1}\otimes\bra{x_2}\right)\ket{\Psi\left(T\right)}
\label{eq:TotalStatePosition}
\end{equation}
of our three coupled system at readout time $T$, for which we still require $T>t_2 >t_1> 0$. Note that this state is again Gaussian, since the interaction, Eq.\,(\ref{eq:InterHam5}), just contains bilinear terms.  However, the interaction will entangle particle and pointers, so that the explicit expression for Eq.\,(\ref{eq:TotalStatePosition}) will be rather lengthy and will not be given here. Nevertheless, we now have a time-evolved state which depends on position variables $x_1$ and $x_2$ of the pointers. Hence we can study various conditioned states of the particle in dependence of specific values $x_i$ found in a single measurement run.

\subsection{Conditioned state via ToF measurements}

To investigate the conditioned state of the particle, we compare probability densities of the particle with and without a measurement of pointer positions. Initially, the probability density of the particle is given by $|\widetilde{\Phi}\left(P\right)|^2$ using Eq.\,(\ref{eq:InitialGaussianParticle}). 

If both pointers have been coupled to the particle, but we do not read off their positions, which means that we trace over the pointer systems, we arrive at the momentum probability density
\begin{align}
W_T\left(P\right)&=\bra{P}\text{tr}_{12}\left(\ket{\Psi\left(T\right)}\bra{\Psi\left(T\right)}\right)\ket{P} \nonumber \\
&=\frac{\exp{\left(-\frac{1}{2}\frac{\left(P-P_0\right)^2}{\Delta P_T^2}\right)}}{\left(2 \pi\Delta P_T^2\right)^{1/2}}
\end{align}
for the particle. The momentum width 
\begin{equation}
\Delta P_T^2=\Delta P_0^2+\kappa^2\left(\Delta p_1^2+\Delta p_2^2\right)\geq\Delta P_0^2
\end{equation} 
corresponds exactly to the variance of the momentum operator, Eq.\,(\ref{eq:HeisenbergMotion15}), in the Heisenberg picture, if we consider an initial state in the form of Eq.\,(\ref{eq:TotalIniStateParticlePointer}). This result clearly indicates the expected disturbance by the pointers: Without observing the coupled pointers we simply add noise to the particle. Furthermore, the mean momentum $P_0$ is not affected by the coupling of the pointers and we see that the limit $\kappa\rightarrow0$ of vanishing coupling leads to the initial momentum distribution $|\widetilde{\Phi}\left(P\right)|^2$.

\begin{figure}[ht]
    \centering
    \includegraphics{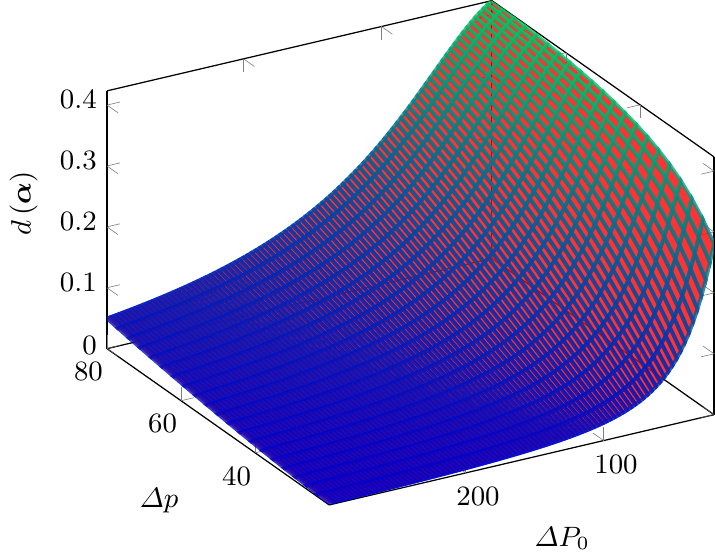}
    \caption[Mean momentum offset]{The gradient $d\left(\boldsymbol{\alpha}\right)$ defined in Eq.\,(\ref{eq:ShiftMean}) as a function of the quantum mechanical widths $\Delta p=\Delta p_1=\Delta p_2$ and $\Delta P_0$. The ToF setup is given by the parameters $\kappa=1$, $t_1=0.5$, $t_2=1.5$, $T=3$, $M/m_1=M/m_2=0.1$ and a mean momentum $P_0=100$ of the particle. In particular we emphasize that this gradient disappears for the initial momentum width $\Delta P_0\rightarrow\infty$ of the particle.}
    \label{fig:Offset}
\end{figure}

The next step is to examine a  truly conditioned probability density of the particle after a specific measurement of the pointer positions $x_i$, i.e. after assigning the ToF value $P_{\mathrm{out}}$, Eq.\,(\ref{eq:OutcomeMomentum}), to the particle. We expect that this measurement leads to a meaningful momentum description for the particle. To confirm this expectation we use Eq.\,(\ref{eq:OutcomeMomentum}) to eliminate $x_2$ in the time-evolved state, Eq.\,(\ref{eq:TotalStatePosition}). Then we trace over the first pointer, renormalize the expression with a constant $\mathcal{N}$ and finally arrive at the conditioned momentum distribution 
\begin{align}
W\left(P|P_{\text{out}}\right)&=\mathcal{N}\!\!\int_{-\infty}^{\infty}{\hspace{-1em}\mathrm{d}x_1\left|\Psi\left(P,x_1,\kappa(t_2-t_1) P_{\mathrm{out}}+x_1,T\right)\right|^2} \nonumber \\
&\equiv\frac{\exp{\left(-\frac{1}{2}\frac{\left(P-P_c\right)^2}{\Delta P_c^2}\right)}}{\left(2\pi\Delta P_c^2\right)^{1/2}}
\label{eq:CondPostMeasDistrib}
\end{align}
of the particle. This momentum distribution is still Gaussian with a conditioned mean $P_c$ and a corresponding uncertainty $\Delta P_c^2$. Ideally, we would expect that $P_c$ coincides with the measured value $P_{\mathrm{out}}$. Actually, in our ToF model we obtain a small linear deviation between the two quantities which can be written as
\begin{equation}
\label{eq:ShiftMean}
	P_c=P_{\mathrm{out}}+d\left(\boldsymbol{\alpha}\right)\left(P_0-P_{\mathrm{out}}\right).
\end{equation} 
The gradient $d\left(\boldsymbol{\alpha}\right)$ depends on all parameters of a specific measurement setup, which means\\ ${\boldsymbol{\alpha}=\left(\kappa, t_1, t_2, T, M/m_1, M/m_2, P_0, \Delta P_0, \Delta p_1, \Delta p_2\right)^T}$.\\ Instead of presenting this involved expression explicitly, we plot the gradient $d\left(\boldsymbol{\alpha}\right)$ in Fig.\,\ref{fig:Offset} as a function of pointer width $\Delta p=\Delta p_1=\Delta p_2$ and initial width $\Delta P_0$ of the measured particle. These parameters actually describe the quantum mechanical content of our ToF scheme. We clearly see that the deviation, Eq.\,(\ref{eq:ShiftMean}), is small and for a particle with broad momentum distribution, which means $\Delta P_0\rightarrow\infty$, it even vanishes. Moreover, we also emphasize that in the statistically most relevant case \footnote{To understand this statistical relevance we recall that due to the equality of mean values, see Eq.\,(\ref{eq:Moment3}), all single ToF values will be centered around $\left<\hat{P}\left(0\right)\right>=P_0$.} we always get $P_{\mathrm{out}}=P_0=P_c$, regardless of all other parameters of the setup.

\begin{figure}
	%\centering
    \subfloat
    {\includegraphics{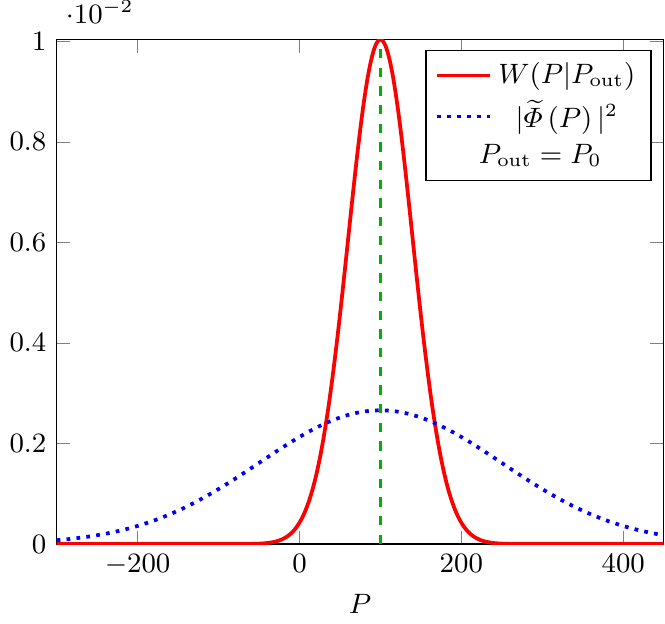}}
    \vspace{-2em}
    \subfloat
    {\includegraphics{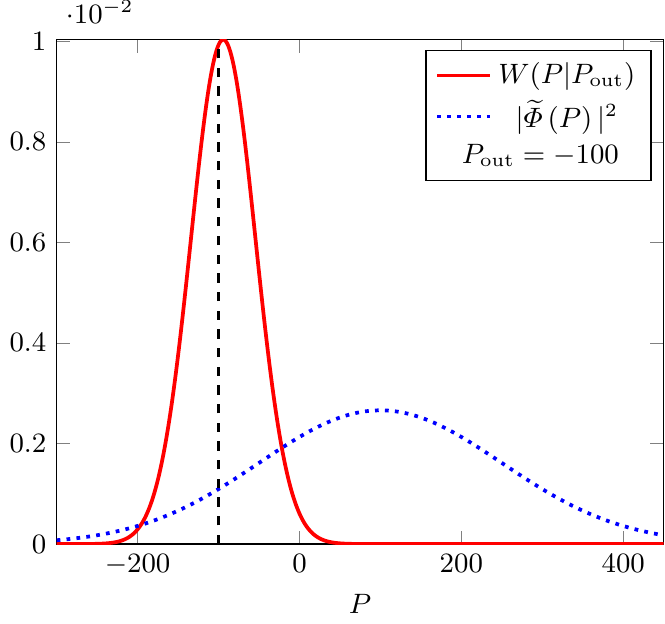}}
    \vspace{-2em}
    \subfloat
    {\includegraphics{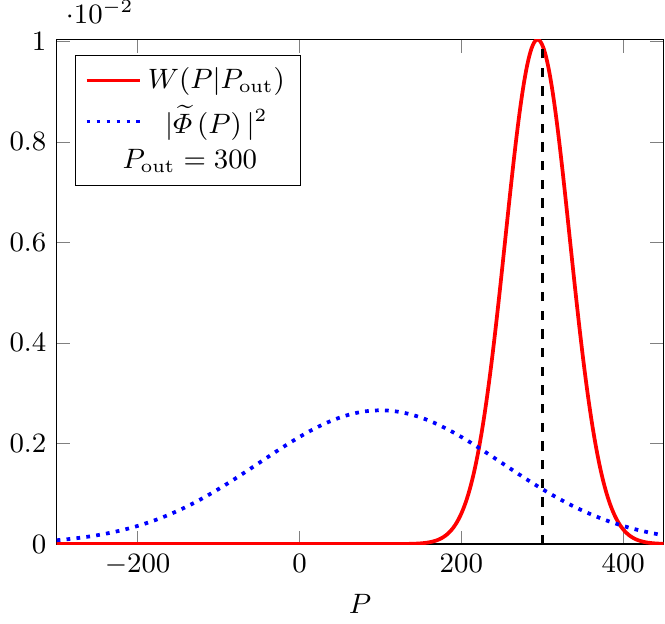}}
    \caption[Probability densities with measurement]{In this figure we exemplify three post-measurement distributions $W\!\left(P|P_{\text{out}}\right)$, Eq.\,(\ref{eq:CondPostMeasDistrib}), and compare them to the initial momentum distribution $|\widetilde{\Phi}\left(P\right)|^2$, Eq.\,(\ref{eq:InitialGaussianParticle}) of the particle. The ToF setup is defined by the parameters already used for Fig.\,\ref{fig:Offset} and we have chosen the widths $\Delta P_0=150$ and $\Delta p=30$. The dashed line now indicates the measured value of $P_{\mathrm{out}}$. We clearly observe that the post-measurement distribution (red solid curve) becomes narrower than the initial distribution (blue dotted curve). This effect is independent of a specific measurement result $P_{\mathrm{out}}$, which might be in the center (a) or in the wings (b) and (c) of the initial distribution. Hence our ToF scheme leads to a \enquote{reduction} of the wave packet in momentum space.}
    \label{fig:ProbDenWithMes}
\end{figure}

It remains to be shown that the width $\Delta P_c$ shrinks under the condition of a specifically measured ToF value $P_{\mathrm{out}}$: Once we acquire the information from the pointers, we reduce the momentum representation for the state of the particle. We illustrate this effect in Fig.\,\ref{fig:ProbDenWithMes} for a set of parameters $\boldsymbol{\alpha}$ which have also been used in Fig.\,\ref{fig:Offset}. 

The conditioned momentum distribution, Eq.\,(\ref{eq:CondPostMeasDistrib}), turns out to possess a considerably reduced width compared to the initial distribution based on Eq.\,(\ref{eq:InitialGaussianParticle}). In fact, this behavior can be seen for a ToF value $P_{\mathrm{out}}=P_0$ as well as for the less probable cases where we find $P_{\mathrm{out}}$ in the wings of the initial distribution. We also note that the small shift $P_c-P_{\mathrm{out}}$, Eq.\,(\ref{eq:ShiftMean}), is hardly visible in Fig.\,\ref{fig:ProbDenWithMes}.

Moreover, we obtain the general result that the post-measurement width $\Delta P_c$ is independent of the initial mean momentum $P_0$ of the particle and the measured momentum $P_{\text{out}}$. This result means that once we have fixed the parameters of the ToF apparatus including the pointers, the post-measurement width $\Delta P_c$ only depends on the initial width $\Delta P_0$ of the particle.

\begin{figure}[ht]
    \centering
    \includegraphics{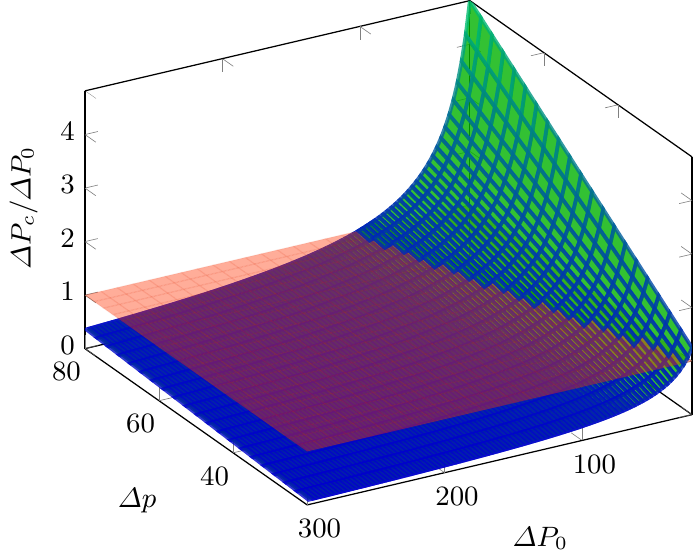}
    \caption[Variance ratio]{In this figure we compare the conditioned post-measurement width $\Delta P_c$ from Eq.\,(\ref{eq:CondPostMeasDistrib}) to the initial width $\Delta P_0$ of the particle. The ToF setup is again defined by the parameters used for Figs.\,\ref{fig:Offset} and \ref{fig:ProbDenWithMes}. Depending on the width $\Delta p$ of the pointers we recognize that the relative post-measurement width $\Delta P_c/\Delta P_0$ only shrinks, that is moves below the indicated critical plane $\Delta P_c/\Delta P_0=1$ (transparent red) if the initial wave packet of the particle is broad enough in momentum space. However, from such a critical value on we clearly discover that a single ToF measurement reduces the wave packet.}
    \label{fig:DiffDeltP}
\end{figure}

However, this dependence has an additional subtlety: By looking at the principle of our ToF scheme depicted in Fig.\,\ref{fig:ClassAndQuantToFTotal} we can understand that the relation $\Delta P_c<\Delta P_0$ will not always be true. An initial wave packet being already quite narrow in momentum space will be spread out over the $x$-axis. Hence, the displacement of each pointer, Eq.\,(\ref{eq:InterHam2}), will fluctuate in a wider range resulting in less accurate information about the particle's momentum downstream of the ToF apparatus. This effect is exactly what we see in Fig.\,\ref{fig:DiffDeltP} where we picture the relative width\\ $\Delta P_c/\Delta P_0$ as a function of the initial width $\Delta P_0$ of the particle and the initial width $\Delta p$ of both pointers. If the initial wave packet reaches a critical uncertainty in momentum, we always observe a conditioned wave packet being narrower: Asymptotically, i.e. $\Delta P_0\rightarrow\infty$, the relative width $\Delta P_c/\Delta P_0$ always vanishes.

\section{Conclusion}
\label{sec:Conclusion}
In this work we have presented an operational quantum model for momentum measurements. It is motivated by classical time-of-flight concepts. Based on a pointer formalism we have defined a ToF observable and then derived a general expression for the mean momentum of the particle. Furthermore, we have examined single momentum measurements based on this ToF notion. We have found that a conditioned post-measurement momentum distribution of the particle will be located around the ToF value. The reduced post-measurement width of the particle has turned out to be independent of the initial mean momentum of the particle and the measurement outcome: It is solely a function of the parameters defining the ToF setup. The process therefore replaces the ad hoc collapse of the particle's state to a momentum eigenstate.

The optimization of the post-measurement width as a function of the initial width of the particle with respect to all parameters of the apparatus including the pointer states will be an interesting starting point for further work. Besides, also the small shift between the post-measured mean momentum and the measurement outcome could be considered for optimization. Another crucial point which has not been treated in this work is entanglement of the pointers: In order to further understand the use of quantum mechanical measurement resources it would be possible to discuss initially entangled pointer states. One can then analyze if these states lead to an even smaller post-measurement width or to a smaller shift between the post-measurement mean momentum and the measurement outcome.

Finally, a promising approach towards an operational description of ToF momentum measurements would be the idea of operational time observables defined and investigated for example in Refs.\,\cite{grot1996time,delgado1997arrival,leon_time--arrival_1997,anastopoulos_time--arrival_2006,anastopoulos_time--arrival_2012,anastopoulos_time--arrival_2017,anastopoulos_time_2019}. Instead of using fixed coupling times as in our model, we could replace them by an operational time observable and analyze if this ansatz still leads to meaningful momentum values and a  state reduction.

\begin{acknowledgements}
	The authors are very grateful to A. Friedrich and E. Giese for discussions, hints and many helpful comments which lead to the finalization of this work. {F. Di Pumpo} thankfully acknowledges the QUANTUS project which is supported by the German Space Agency DLR with funds provided by the Federal Ministry of Economics and Technology (BMWi) under grant number 50WM1556 (QUANTUS IV).
	
	This is a pre-print of an article published in the \textit{European Physical Journal}. The final authenticated version is available online at: \url{https://doi.org/10.1140/epjd/e2019-100226-1}.
\end{acknowledgements}

\section*{Authors contributions}
Both authors were involved in the preparation of the\\ manuscript and both authors have read and approved the final manuscript.

%\bibliography{Literatur}

\begin{thebibliography}{37}%
	\makeatletter
	\providecommand \@ifxundefined [1]{%
		\@ifx{#1\undefined}
	}%
	\providecommand \@ifnum [1]{%
		\ifnum #1\expandafter \@firstoftwo
		\else \expandafter \@secondoftwo
		\fi
	}%
	\providecommand \@ifx [1]{%
		\ifx #1\expandafter \@firstoftwo
		\else \expandafter \@secondoftwo
		\fi
	}%
	\providecommand \natexlab [1]{#1}%
	\providecommand \enquote  [1]{``#1''}%
	\providecommand \bibnamefont  [1]{#1}%
	\providecommand \bibfnamefont [1]{#1}%
	\providecommand \citenamefont [1]{#1}%
	\providecommand \href@noop [0]{\@secondoftwo}%
	\providecommand \href [0]{\begingroup \@sanitize@url \@href}%
	\providecommand \@href[1]{\@@startlink{#1}\@@href}%
	\providecommand \@@href[1]{\endgroup#1\@@endlink}%
	\providecommand \@sanitize@url [0]{\catcode `\\12\catcode `\$12\catcode
		`\&12\catcode `\#12\catcode `\^12\catcode `\_12\catcode `\%12\relax}%
	\providecommand \@@startlink[1]{}%
	\providecommand \@@endlink[0]{}%
	\providecommand \url  [0]{\begingroup\@sanitize@url \@url }%
	\providecommand \@url [1]{\endgroup\@href {#1}{\urlprefix }}%
	\providecommand \urlprefix  [0]{URL }%
	\providecommand \Eprint [0]{\href }%
	\providecommand \doibase [0]{https://doi.org/}%
	\providecommand \selectlanguage [0]{\@gobble}%
	\providecommand \bibinfo  [0]{\@secondoftwo}%
	\providecommand \bibfield  [0]{\@secondoftwo}%
	\providecommand \translation [1]{[#1]}%
	\providecommand \BibitemOpen [0]{}%
	\providecommand \bibitemStop [0]{}%
	\providecommand \bibitemNoStop [0]{.\EOS\space}%
	\providecommand \EOS [0]{\spacefactor3000\relax}%
	\providecommand \BibitemShut  [1]{\csname bibitem#1\endcsname}%
	\let\auto@bib@innerbib\@empty
	%</preamble>
	\bibitem [{\citenamefont {von
			Neumann}(1932)}]{VonNeumann1968MathematischeGrundlagen}%
	\BibitemOpen
	\bibfield  {author} {\bibinfo {author} {\bibfnamefont {J.}~\bibnamefont {von
				Neumann}},\ }\href@noop {} {\emph {\bibinfo {title} {Mathematische Grundlagen
				der Quantenmechanik}}}\ (\bibinfo  {publisher} {Springer Berlin Heidelberg},\
	\bibinfo {year} {1932})\BibitemShut {NoStop}%
	\bibitem [{\citenamefont {Gallone}(2014)}]{gallone2014hilbert}%
	\BibitemOpen
	\bibfield  {author} {\bibinfo {author} {\bibfnamefont {F.}~\bibnamefont
			{Gallone}},\ }\href@noop {} {\emph {\bibinfo {title} {Hilbert Space and
				Quantum Mechanics}}}\ (\bibinfo  {publisher} {World Scientific Publishing
		Company},\ \bibinfo {year} {2014})\BibitemShut {NoStop}%
	\bibitem [{\citenamefont {David}(2014)}]{david2014formalisms}%
	\BibitemOpen
	\bibfield  {author} {\bibinfo {author} {\bibfnamefont {F.}~\bibnamefont
			{David}},\ }\href@noop {} {\emph {\bibinfo {title} {The Formalisms of Quantum
				Mechanics: An Introduction}}},\ Lecture Notes in Physics\ (\bibinfo
	{publisher} {Springer International Publishing},\ \bibinfo {year}
	{2014})\BibitemShut {NoStop}%
	\bibitem [{\citenamefont {Lamb}(1969)}]{lamb1969operational}%
	\BibitemOpen
	\bibfield  {author} {\bibinfo {author} {\bibfnamefont {W.~E.}\ \bibnamefont
			{Lamb}},\ }\bibfield  {title} {\bibinfo {title} {An operational
			interpretation of nonrelativistic quantum mechanics},\ }\href
	{http://physicstoday.scitation.org/doi/10.1063/1.3035523} {\bibfield
		{journal} {\bibinfo  {journal} {Phys. Today}\ }\textbf {\bibinfo {volume} {22
				(4)}},\ \bibinfo {pages} {23} (\bibinfo {year} {1969})}\BibitemShut {NoStop}%
	\bibitem [{\citenamefont {Zurek}(1981)}]{PhysRevD.24.1516}%
	\BibitemOpen
	\bibfield  {author} {\bibinfo {author} {\bibfnamefont {W.~H.}\ \bibnamefont
			{Zurek}},\ }\bibfield  {title} {\bibinfo {title} {Pointer basis of quantum
			apparatus: Into what mixture does the wave packet collapse?},\ }\href
	{https://link.aps.org/doi/10.1103/PhysRevD.24.1516} {\bibfield  {journal}
		{\bibinfo  {journal} {Phys. Rev. D}\ }\textbf {\bibinfo {volume} {24}},\
		\bibinfo {pages} {1516} (\bibinfo {year} {1981})}\BibitemShut {NoStop}%
	\bibitem [{\citenamefont {Zurek}(1982)}]{PhysRevD.26.1862}%
	\BibitemOpen
	\bibfield  {author} {\bibinfo {author} {\bibfnamefont {W.~H.}\ \bibnamefont
			{Zurek}},\ }\bibfield  {title} {\bibinfo {title} {Environment-induced
			superselection rules},\ }\href
	{https://link.aps.org/doi/10.1103/PhysRevD.26.1862} {\bibfield  {journal}
		{\bibinfo  {journal} {Phys. Rev. D}\ }\textbf {\bibinfo {volume} {26}},\
		\bibinfo {pages} {1862} (\bibinfo {year} {1982})}\BibitemShut {NoStop}%
	\bibitem [{\citenamefont {Busch}\ \emph {et~al.}(1995)\citenamefont {Busch},
		\citenamefont {Grabowski},\ and\ \citenamefont
		{Lahti}}]{busch1997operational}%
	\BibitemOpen
	\bibfield  {author} {\bibinfo {author} {\bibfnamefont {P.}~\bibnamefont
			{Busch}}, \bibinfo {author} {\bibfnamefont {M.}~\bibnamefont {Grabowski}},\
		and\ \bibinfo {author} {\bibfnamefont {P.~J.}\ \bibnamefont {Lahti}},\
	}\href@noop {} {\emph {\bibinfo {title} {Operational {Q}uantum {P}hysics}}}\
	(\bibinfo  {publisher} {Springer Berlin Heidelberg},\ \bibinfo {year}
	{1995})\BibitemShut {NoStop}%
	\bibitem [{\citenamefont {Klyshko}(1998)}]{klyshko1998basic}%
	\BibitemOpen
	\bibfield  {author} {\bibinfo {author} {\bibfnamefont {D.~N.}\ \bibnamefont
			{Klyshko}},\ }\bibfield  {title} {\bibinfo {title} {Basic quantum mechanical
			concepts from the operational viewpoint},\ }\href
	{http://stacks.iop.org/1063-7869/41/i=9/a=R05?key=crossref.68fb04de86908a689ab8f11f25fdc32e}
	{\bibfield  {journal} {\bibinfo  {journal} {Phys.-Uspekhi}\ }\textbf
		{\bibinfo {volume} {41}},\ \bibinfo {pages} {885} (\bibinfo {year}
		{1998})}\BibitemShut {NoStop}%
	\bibitem [{\citenamefont {D’Ariano}(2007)}]{dariano2007Operational}%
	\BibitemOpen
	\bibfield  {author} {\bibinfo {author} {\bibfnamefont {G.~M.}\ \bibnamefont
			{D’Ariano}},\ }\bibfield  {title} {\bibinfo {title} {Operational {A}xioms
			for {Q}uantum {M}echanics},\ }\href
	{https://aip.scitation.org/doi/abs/10.1063/1.2713449} {\bibfield  {journal}
		{\bibinfo  {journal} {AIP Conf. Proc.}\ }\textbf {\bibinfo {volume} {889}},\
		\bibinfo {pages} {79} (\bibinfo {year} {2007})}\BibitemShut {NoStop}%
	\bibitem [{\citenamefont {Englert}\ and\ \citenamefont
		{W\'odkiewicz}(1995)}]{PhysRevA.51.R2661}%
	\BibitemOpen
	\bibfield  {author} {\bibinfo {author} {\bibfnamefont {B.-G.}\ \bibnamefont
			{Englert}}\ and\ \bibinfo {author} {\bibfnamefont {K.}~\bibnamefont
			{W\'odkiewicz}},\ }\bibfield  {title} {\bibinfo {title} {Intrinsic and
			operational observables in quantum mechanics},\ }\href
	{https://link.aps.org/doi/10.1103/PhysRevA.51.R2661} {\bibfield  {journal}
		{\bibinfo  {journal} {Phys. Rev. A}\ }\textbf {\bibinfo {volume} {51}},\
		\bibinfo {pages} {R2661} (\bibinfo {year} {1995})}\BibitemShut {NoStop}%
	\bibitem [{\citenamefont {Hillery}\ \emph {et~al.}(1984)\citenamefont
		{Hillery}, \citenamefont {O'Connell}, \citenamefont {Scully},\ and\
		\citenamefont {Wigner}}]{HILLERY1984121}%
	\BibitemOpen
	\bibfield  {author} {\bibinfo {author} {\bibfnamefont {M.}~\bibnamefont
			{Hillery}}, \bibinfo {author} {\bibfnamefont {R.~F.}\ \bibnamefont
			{O'Connell}}, \bibinfo {author} {\bibfnamefont {M.~O.}\ \bibnamefont
			{Scully}},\ and\ \bibinfo {author} {\bibfnamefont {E.~P.}\ \bibnamefont
			{Wigner}},\ }\bibfield  {title} {\bibinfo {title} {Distribution functions in
			physics: Fundamentals},\ }\href
	{http://www.sciencedirect.com/science/article/pii/0370157384901601}
	{\bibfield  {journal} {\bibinfo  {journal} {Phys. Rep.}\ }\textbf {\bibinfo
			{volume} {106}},\ \bibinfo {pages} {121 } (\bibinfo {year}
		{1984})}\BibitemShut {NoStop}%
	\bibitem [{\citenamefont {Schleich}(2001)}]{schleich2001quantum}%
	\BibitemOpen
	\bibfield  {author} {\bibinfo {author} {\bibfnamefont {W.~P.}\ \bibnamefont
			{Schleich}},\ }\href@noop {} {\emph {\bibinfo {title} {Quantum {O}ptics in
				{P}hase {S}pace}}}\ (\bibinfo  {publisher} {Wiley-VCH Berlin Weinheim},\
	\bibinfo {year} {2001})\BibitemShut {NoStop}%
	\bibitem [{\citenamefont {Vogel}\ and\ \citenamefont
		{Risken}(1989)}]{vogel1989determination}%
	\BibitemOpen
	\bibfield  {author} {\bibinfo {author} {\bibfnamefont {K.}~\bibnamefont
			{Vogel}}\ and\ \bibinfo {author} {\bibfnamefont {H.}~\bibnamefont {Risken}},\
	}\bibfield  {title} {\bibinfo {title} {Determination of quasiprobability
			distributions in terms of probability distributions for the rotated
			quadrature phase},\ }\href {https://dx.doi.org/10.1103/PhysRevA.40.2847}
	{\bibfield  {journal} {\bibinfo  {journal} {Phys. Rev. A}\ }\textbf {\bibinfo
			{volume} {40}},\ \bibinfo {pages} {2847} (\bibinfo {year}
		{1989})}\BibitemShut {NoStop}%
	\bibitem [{\citenamefont {Leonhardt}(1997)}]{leonhardt1997measuring}%
	\BibitemOpen
	\bibfield  {author} {\bibinfo {author} {\bibfnamefont {U.}~\bibnamefont
			{Leonhardt}},\ }\href@noop {} {\emph {\bibinfo {title} {Measuring the
				{Q}uantum {S}tate of {L}ight}}},\ Vol.~\bibinfo {volume} {22}\ (\bibinfo
	{publisher} {Cambridge {U}niversity {P}ress},\ \bibinfo {year}
	{1997})\BibitemShut {NoStop}%
	\bibitem [{\citenamefont {Paris}\ and\ \citenamefont
		{Rehacek}(2004)}]{paris2004quantum}%
	\BibitemOpen
	\bibfield  {author} {\bibinfo {author} {\bibfnamefont {M.}~\bibnamefont
			{Paris}}\ and\ \bibinfo {author} {\bibfnamefont {J.}~\bibnamefont
			{Rehacek}},\ }\href@noop {} {\emph {\bibinfo {title} {Quantum State
				Estimation}}},\ Lecture Notes in Physics\ (\bibinfo  {publisher} {Springer
		Berlin Heidelberg},\ \bibinfo {year} {2004})\BibitemShut {NoStop}%
	\bibitem [{\citenamefont {Arthurs}\ and\ \citenamefont
		{Kelly}(1965)}]{BLTJ:BLTJ1684}%
	\BibitemOpen
	\bibfield  {author} {\bibinfo {author} {\bibfnamefont {E.}~\bibnamefont
			{Arthurs}}\ and\ \bibinfo {author} {\bibfnamefont {J.~L.}\ \bibnamefont
			{Kelly}},\ }\bibfield  {title} {\bibinfo {title} {On the {S}imultaneous
			{M}easurement of a {P}air of {C}onjugate {O}bservables},\ }\href
	{https://dx.doi.org/10.1002/j.1538-7305.1965.tb01684.x} {\bibfield  {journal}
		{\bibinfo  {journal} {Bell Syst. Tech. J.}\ }\textbf {\bibinfo {volume}
			{44}},\ \bibinfo {pages} {725} (\bibinfo {year} {1965})}\BibitemShut
	{NoStop}%
	\bibitem [{\citenamefont {Park}\ and\ \citenamefont
		{Margenau}(1968)}]{park_simultaneous_1968}%
	\BibitemOpen
	\bibfield  {author} {\bibinfo {author} {\bibfnamefont {J.~L.}\ \bibnamefont
			{Park}}\ and\ \bibinfo {author} {\bibfnamefont {H.}~\bibnamefont
			{Margenau}},\ }\bibfield  {title} {\bibinfo {title} {Simultaneous
			measurability in quantum theory},\ }\href
	{https://dx.doi.org/10.1007/BF00668668} {\bibfield  {journal} {\bibinfo
			{journal} {Int. J. Theor. Phys.}\ }\textbf {\bibinfo {volume} {1}},\ \bibinfo
		{pages} {211} (\bibinfo {year} {1968})}\BibitemShut {NoStop}%
	\bibitem [{\citenamefont {Stenholm}(1992)}]{STENHOLM1992233}%
	\BibitemOpen
	\bibfield  {author} {\bibinfo {author} {\bibfnamefont {S.}~\bibnamefont
			{Stenholm}},\ }\bibfield  {title} {\bibinfo {title} {Simultaneous measurement
			of conjugate variables},\ }\href
	{http://www.sciencedirect.com/science/article/pii/0003491692900862}
	{\bibfield  {journal} {\bibinfo  {journal} {Ann. Phys.}\ }\textbf {\bibinfo
			{volume} {218}},\ \bibinfo {pages} {233 } (\bibinfo {year}
		{1992})}\BibitemShut {NoStop}%
	\bibitem [{\citenamefont {Busch}(1985)}]{busch1985indeterminacy}%
	\BibitemOpen
	\bibfield  {author} {\bibinfo {author} {\bibfnamefont {P.}~\bibnamefont
			{Busch}},\ }\bibfield  {title} {\bibinfo {title} {Indeterminacy relations and
			simultaneous measurements in quantum theory},\ }\href
	{http://link.springer.com/10.1007/BF00670074} {\bibfield  {journal} {\bibinfo
			{journal} {Int. J. Theor. Phys.}\ }\textbf {\bibinfo {volume} {24}},\
		\bibinfo {pages} {63} (\bibinfo {year} {1985})}\BibitemShut {NoStop}%
	\bibitem [{\citenamefont {Wódkiewicz}(1987)}]{WODKIEWICZ1987207}%
	\BibitemOpen
	\bibfield  {author} {\bibinfo {author} {\bibfnamefont {K.}~\bibnamefont
			{Wódkiewicz}},\ }\bibfield  {title} {\bibinfo {title} {On the operational
			uncertainty relation},\ }\href
	{http://www.sciencedirect.com/science/article/pii/0375960187906219}
	{\bibfield  {journal} {\bibinfo  {journal} {Phys. Lett. A}\ }\textbf
		{\bibinfo {volume} {124}},\ \bibinfo {pages} {207 } (\bibinfo {year}
		{1987})}\BibitemShut {NoStop}%
	\bibitem [{\citenamefont {Bu{\ss}hardt}\ and\ \citenamefont
		{Freyberger}(2010)}]{busshardt2010timing}%
	\BibitemOpen
	\bibfield  {author} {\bibinfo {author} {\bibfnamefont {M.}~\bibnamefont
			{Bu{\ss}hardt}}\ and\ \bibinfo {author} {\bibfnamefont {M.}~\bibnamefont
			{Freyberger}},\ }\bibfield  {title} {\bibinfo {title} {Timing in quantum
			measurements of position and momentum},\ }\href
	{https://link.aps.org/doi/10.1103/PhysRevA.82.042117} {\bibfield  {journal}
		{\bibinfo  {journal} {Phys. Rev. A}\ }\textbf {\bibinfo {volume} {82}},\
		\bibinfo {pages} {042117} (\bibinfo {year} {2010})}\BibitemShut {NoStop}%
	\bibitem [{\citenamefont {Heese}\ and\ \citenamefont
		{Freyberger}(2013)}]{heese2013entropic}%
	\BibitemOpen
	\bibfield  {author} {\bibinfo {author} {\bibfnamefont {R.}~\bibnamefont
			{Heese}}\ and\ \bibinfo {author} {\bibfnamefont {M.}~\bibnamefont
			{Freyberger}},\ }\bibfield  {title} {\bibinfo {title} {Entropic uncertainty
			relation for pointer-based simultaneous measurements of conjugate
			observables},\ }\href {https://link.aps.org/doi/10.1103/PhysRevA.87.012123}
	{\bibfield  {journal} {\bibinfo  {journal} {Phys. Rev. A}\ }\textbf {\bibinfo
			{volume} {87}},\ \bibinfo {pages} {012123} (\bibinfo {year}
		{2013})}\BibitemShut {NoStop}%
	\bibitem [{\citenamefont {{For a review see: J. G. Muga and C. R.
				Leavens}}(2000)}]{muga2000arrival}%
	\BibitemOpen
	\bibfield  {author} {\bibinfo {author} {\bibnamefont {{For a review see: J.
					G. Muga and C. R. Leavens}}},\ }\bibfield  {title} {\bibinfo {title} {Arrival
			time in quantum mechanics},\ }\href
	{https://dx.doi.org/10.1016/S0370-1573(00)00047-8} {\bibfield  {journal}
		{\bibinfo  {journal} {Phys. Rep.}\ }\textbf {\bibinfo {volume} {338}},\
		\bibinfo {pages} {353} (\bibinfo {year} {2000})}\BibitemShut {NoStop}%
	\bibitem [{\citenamefont {Grot}\ \emph {et~al.}(1996)\citenamefont {Grot},
		\citenamefont {Rovelli},\ and\ \citenamefont {Tate}}]{grot1996time}%
	\BibitemOpen
	\bibfield  {author} {\bibinfo {author} {\bibfnamefont {N.}~\bibnamefont
			{Grot}}, \bibinfo {author} {\bibfnamefont {C.}~\bibnamefont {Rovelli}},\ and\
		\bibinfo {author} {\bibfnamefont {R.~S.}\ \bibnamefont {Tate}},\ }\bibfield
	{title} {\bibinfo {title} {Time of arrival in quantum mechanics},\ }\href
	{https://link.aps.org/doi/10.1103/PhysRevA.54.4676} {\bibfield  {journal}
		{\bibinfo  {journal} {Phys. Rev. A}\ }\textbf {\bibinfo {volume} {54}},\
		\bibinfo {pages} {4676} (\bibinfo {year} {1996})}\BibitemShut {NoStop}%
	\bibitem [{\citenamefont {Delgado}\ and\ \citenamefont
		{Muga}(1997)}]{delgado1997arrival}%
	\BibitemOpen
	\bibfield  {author} {\bibinfo {author} {\bibfnamefont {V.}~\bibnamefont
			{Delgado}}\ and\ \bibinfo {author} {\bibfnamefont {J.~G.}\ \bibnamefont
			{Muga}},\ }\bibfield  {title} {\bibinfo {title} {Arrival time in quantum
			mechanics},\ }\href {https://link.aps.org/doi/10.1103/PhysRevA.56.3425}
	{\bibfield  {journal} {\bibinfo  {journal} {Phys. Rev. A}\ }\textbf {\bibinfo
			{volume} {56}},\ \bibinfo {pages} {3425} (\bibinfo {year}
		{1997})}\BibitemShut {NoStop}%
	\bibitem [{\citenamefont {León}(1997)}]{leon_time--arrival_1997}%
	\BibitemOpen
	\bibfield  {author} {\bibinfo {author} {\bibfnamefont {J.}~\bibnamefont
			{León}},\ }\bibfield  {title} {\bibinfo {title} {Time-of-arrival formalism
			for the relativistic particle},\ }\href
	{http://stacks.iop.org/0305-4470/30/i=13/a=027?key=crossref.d9ae1dda3a38ef1e1a04f5bccc70d195}
	{\bibfield  {journal} {\bibinfo  {journal} {J. Phys. A}\ }\textbf {\bibinfo
			{volume} {30}},\ \bibinfo {pages} {4791} (\bibinfo {year}
		{1997})}\BibitemShut {NoStop}%
	\bibitem [{\citenamefont {Anastopoulos}\ and\ \citenamefont
		{Savvidou}(2006)}]{anastopoulos_time--arrival_2006}%
	\BibitemOpen
	\bibfield  {author} {\bibinfo {author} {\bibfnamefont {C.}~\bibnamefont
			{Anastopoulos}}\ and\ \bibinfo {author} {\bibfnamefont {N.}~\bibnamefont
			{Savvidou}},\ }\bibfield  {title} {\bibinfo {title} {Time-of-arrival
			probabilities and quantum measurements},\ }\href
	{http://aip.scitation.org/doi/10.1063/1.2399085} {\bibfield  {journal}
		{\bibinfo  {journal} {J. Math. Phys.}\ }\textbf {\bibinfo {volume} {47}},\
		\bibinfo {pages} {122106} (\bibinfo {year} {2006})}\BibitemShut {NoStop}%
	\bibitem [{\citenamefont {Anastopoulos}\ and\ \citenamefont
		{Savvidou}(2012)}]{anastopoulos_time--arrival_2012}%
	\BibitemOpen
	\bibfield  {author} {\bibinfo {author} {\bibfnamefont {C.}~\bibnamefont
			{Anastopoulos}}\ and\ \bibinfo {author} {\bibfnamefont {N.}~\bibnamefont
			{Savvidou}},\ }\bibfield  {title} {\bibinfo {title} {Time-of-arrival
			probabilities for general particle detectors},\ }\href
	{https://link.aps.org/doi/10.1103/PhysRevA.86.012111} {\bibfield  {journal}
		{\bibinfo  {journal} {Phys. Rev. A}\ }\textbf {\bibinfo {volume} {86}}
		(\bibinfo {year} {2012})}\BibitemShut {NoStop}%
	\bibitem [{\citenamefont {Anastopoulos}\ and\ \citenamefont
		{Savvidou}(2017)}]{anastopoulos_time--arrival_2017}%
	\BibitemOpen
	\bibfield  {author} {\bibinfo {author} {\bibfnamefont {C.}~\bibnamefont
			{Anastopoulos}}\ and\ \bibinfo {author} {\bibfnamefont {N.}~\bibnamefont
			{Savvidou}},\ }\bibfield  {title} {\bibinfo {title} {Time-of-arrival
			correlations},\ }\href {https://link.aps.org/doi/10.1103/PhysRevA.95.032105}
	{\bibfield  {journal} {\bibinfo  {journal} {Phys. Rev. A}\ }\textbf {\bibinfo
			{volume} {95}} (\bibinfo {year} {2017})}\BibitemShut {NoStop}%
	\bibitem [{\citenamefont {Anastopoulos}\ and\ \citenamefont
		{Savvidou}(2019)}]{anastopoulos_time_2019}%
	\BibitemOpen
	\bibfield  {author} {\bibinfo {author} {\bibfnamefont {C.}~\bibnamefont
			{Anastopoulos}}\ and\ \bibinfo {author} {\bibfnamefont {N.}~\bibnamefont
			{Savvidou}},\ }\bibfield  {title} {\bibinfo {title} {Time of arrival and
			localization of relativistic particles},\ }\href
	{https://doi.org/10.1063/1.5080930} {\bibfield  {journal} {\bibinfo
			{journal} {J. Math. Phys.}\ }\textbf {\bibinfo {volume}
			{60}},\ \bibinfo {pages} {032301} (\bibinfo {year} {2019})}\BibitemShut
	{NoStop}%
	\bibitem [{\citenamefont {Aharonov}\ and\ \citenamefont
		{Bohm}(1961)}]{aharonov_time1961}%
	\BibitemOpen
	\bibfield  {author} {\bibinfo {author} {\bibfnamefont {Y.}~\bibnamefont
			{Aharonov}}\ and\ \bibinfo {author} {\bibfnamefont {D.}~\bibnamefont
			{Bohm}},\ }\bibfield  {title} {\bibinfo {title} {Time in the {Quantum}
			{Theory} and the {Uncertainty} {Relation} for {Time} and {Energy}},\ }\href
	{https://link.aps.org/doi/10.1103/PhysRev.122.1649} {\bibfield  {journal}
		{\bibinfo  {journal} {Phys. Rev.}\ }\textbf {\bibinfo {volume} {122}},\
		\bibinfo {pages} {1649} (\bibinfo {year} {1961})}\BibitemShut {NoStop}%
	\bibitem [{\citenamefont {Kijowski}(1974)}]{kijowski_time1974}%
	\BibitemOpen
	\bibfield  {author} {\bibinfo {author} {\bibfnamefont {J.}~\bibnamefont
			{Kijowski}},\ }\bibfield  {title} {\bibinfo {title} {On the time operator in
			quantum mechanics and the heisenberg uncertainty relation for energy and
			time},\ }\href
	{http://linkinghub.elsevier.com/retrieve/pii/S0034487774800042} {\bibfield
		{journal} {\bibinfo  {journal} {Rep. Math. Phys.}\ }\textbf {\bibinfo
			{volume} {6}},\ \bibinfo {pages} {361} (\bibinfo {year} {1974})}\BibitemShut
	{NoStop}%
	\bibitem [{\citenamefont {Peres}(1980)}]{peres_measurement1980}%
	\BibitemOpen
	\bibfield  {author} {\bibinfo {author} {\bibfnamefont {A.}~\bibnamefont
			{Peres}},\ }\bibfield  {title} {\bibinfo {title} {Measurement of time by
			quantum clocks},\ }\href {http://aapt.scitation.org/doi/10.1119/1.12061}
	{\bibfield  {journal} {\bibinfo  {journal} {Am. J. Phys.}\ }\textbf {\bibinfo
			{volume} {48}},\ \bibinfo {pages} {552} (\bibinfo {year} {1980})}\BibitemShut
	{NoStop}%
	\bibitem [{\citenamefont {Aharonov}\ \emph {et~al.}(1998)\citenamefont
		{Aharonov}, \citenamefont {Oppenheim}, \citenamefont {Popescu}, \citenamefont
		{Reznik},\ and\ \citenamefont {Unruh}}]{aharonov_measurement1998}%
	\BibitemOpen
	\bibfield  {author} {\bibinfo {author} {\bibfnamefont {Y.}~\bibnamefont
			{Aharonov}}, \bibinfo {author} {\bibfnamefont {J.}~\bibnamefont {Oppenheim}},
		\bibinfo {author} {\bibfnamefont {S.}~\bibnamefont {Popescu}}, \bibinfo
		{author} {\bibfnamefont {B.}~\bibnamefont {Reznik}},\ and\ \bibinfo {author}
		{\bibfnamefont {W.~G.}\ \bibnamefont {Unruh}},\ }\bibfield  {title} {\bibinfo
		{title} {Measurement of time of arrival in quantum mechanics},\ }\href
	{https://link.aps.org/doi/10.1103/PhysRevA.57.4130} {\bibfield  {journal}
		{\bibinfo  {journal} {Phys. Rev. A}\ }\textbf {\bibinfo {volume} {57}},\
		\bibinfo {pages} {4130} (\bibinfo {year} {1998})}\BibitemShut {NoStop}%
	\bibitem [{\citenamefont {Yearsley}\ \emph {et~al.}(2011)\citenamefont
		{Yearsley}, \citenamefont {Downs}, \citenamefont {Halliwell},\ and\
		\citenamefont {Hashagen}}]{yearsley_quantum2011}%
	\BibitemOpen
	\bibfield  {author} {\bibinfo {author} {\bibfnamefont {J.~M.}\ \bibnamefont
			{Yearsley}}, \bibinfo {author} {\bibfnamefont {D.~A.}\ \bibnamefont {Downs}},
		\bibinfo {author} {\bibfnamefont {J.~J.}\ \bibnamefont {Halliwell}},\ and\
		\bibinfo {author} {\bibfnamefont {A.~K.}\ \bibnamefont {Hashagen}},\
	}\bibfield  {title} {\bibinfo {title} {Quantum arrival and dwell times via
			idealized clocks},\ }\href
	{https://link.aps.org/doi/10.1103/PhysRevA.84.022109} {\bibfield  {journal}
		{\bibinfo  {journal} {Phys. Rev. A}\ }\textbf {\bibinfo {volume} {84}}
		(\bibinfo {year} {2011})}\BibitemShut {NoStop}%
	\bibitem [{Note1()}]{Note1}%
	\BibitemOpen
	\bibinfo {note} {This quantum scenario of a ToF measurement is closest to the
		classical scenario of taking snapshots on a movie with a camera of the
		classical particle traveling along the $x$-axis.}\BibitemShut {Stop}%
	\bibitem [{Note2()}]{Note2}%
	\BibitemOpen
	\bibinfo {note} {To understand this statistical relevance we recall that due
		to the equality of mean values, see Eq.\protect \tmspace +\thinmuskip
		{.1667em}(\ref {eq:Moment3}), all single ToF values will be centered around
		$\left <\protect \mathaccentV {hat}05E{P}\left (0\right )\right
		>=P_0$.}\BibitemShut {Stop}%
\end{thebibliography}

\end{document}